# A Novel Chaos-Based Cryptographic Scrambling Technique to Secure Medical Images

Chandra Sekhar Sanaboina

Assistant Professor, Department of Computer Science and Engineering, University College of Engineering Ka-kinada(A),
Jawaharlal Nehru Technological University Kakinada, Andhra Pradesh, India
Email address: chandrasekhar.s@jntucek.ac.in

*Abstract*— These days, a tremendous quantity of digital visual data is sent over many networks and stored in many different formats. This visual information is usually very confidential and financially rewarding. Maintaining safe trans-mission of data is crucial, as is the use of approaches to offer security features like privacy, integrity, or authenti-cation that are tailored to certain types of data. Protecting sensitive medical images stored in electronic health records is the focus of this article, which proposes a technique of encryption and decryption. In order to safe-guard image-based programs, encryption methods are applied. Privacy, integrity, and authenticity are only few of the security elements investigated by the proposed system, which encrypts medical pictures using chaos maps. In all stages of the protocol, the suggested chaos-based data scrambling method is employed to mitigate the short-comings of traditional confusion and diffusion designs. Bifurcation charts, Lyapunov exponents, tests for mean squared error and peak-to-average signal-to-noise ratio, and histogram analysis are only some of the tools we use to investigate the suggested system's chaotic behavior.

## I. INTRODUCTION

Today is the dawn of the information age. It is crucial to archive data from every area of our existence. That being the case, data is a commodity in its own right. The data is a valuable asset that must be protected. The three pillars of information security—confidentiality, integrity, and availability—ensure that data is protected from prying eyes (confidentiality), is not tampered with (integrity), and is only made available (availability) when absolutely required ([1]). It is important to protect data not just when it is being inputted into a computer but also while it is being transported over the internet or other public networks.

*Introduction to Telemedicine*

Medical imaging technologies such as X-ray, ultrasound, MRI, and internal computed tomography (CT) provide the two- and three-dimensional images used extensively in contemporary digital health care [2, 3]. According to [3], including X-rays, computed X-ray imaging, magnetic fields, magnetic field gradients, electron microscopy, X-ray imaging, and ultrasound, there are a number of physical techniques that may be used to create these images. The inability of physicians to see one other's patients' records is a common problem in today's healthcare systems. Preventing resource waste and ensuring that medical reports for each patient are written in the correct chronological order, repeated collections of almost identical medical imaging data must be avoided, as must the loss of any such data that has already been collected.

A network of distributed databases [4] might be one solution to these problems, since it would enable all doctors to have electronic access to their patients' preexisting medical imaging data, in particular any long-term medical imaging data collection. The need of safeguarding individual patients' diagnostic imaging records stored in databases or sent across networks is underscored by these numbers. These photos may include sensitive patient information that might be compromised if they are used inappropriately [5].

*Chaos and Cryptography*

Most people are familiar with two hallmarks of chaos: the "butterfly effect," or sensitivity to initial circumstances, as shown in [6], and the pseudo-randomness created by deterministic equations, which is discussed in [7]. Traditional cryptosystems have many key traits with chaotic systems, including ergodicity, sensitivity to initial conditions and control settings, and unpredictable behavior. A deterministic chaotic system is one whose behavior can be predicted using known equations. Since the results of applying the chaos map repeatedly to the same two locations will be sensitive to the initial conditions, the iterations will ultimately diverge and lose their connection [8]. Therefore, chaotic systems may be used as artificial RNGs. Each route in the chaotic map will go through every area of the state space, regardless of how many regions there are, since the map is ergodic. Ergodicity is guaranteed for topologically chaotic maps. The central idea is that, as shown in [5, 6], [8], there are similarities between (continuous) chaotic maps and (discrete) cryptosystems.

## II. RELATED WORK

It has become more important in recent years to guarantee secure picture transmission as digital image sharing has grown more commonplace. Image encryption has been incorporated to provide safety for photo apps. It is a summary of the different picture encryption techniques in use today. Some characteristics of efficient photo encoding techniques are also outlined.

In ref. [7], I. Yasser, A. T. Khalil, M. A. Mohamed, A. S. Samra, and F. Khalifa investigate new chaotic maps and innovative perturbation techniques. Pixels' randomization and replacement are controlled by the resultant characteristics, while the scheme's permutation and diffusion are managed by chaotic sequences and revised mapping parameters. A innovative approach for encrypting grayscale and color medium pictures based on chaos and picture blocks is described by S. T. Kamal, K. M. Hosny, T. M. Elgindy, M. M. Darwish, and M. M. Fouda in ref. [9]. Using image blocks is an innovative method of segmentation. When it comes to protecting sensitive medical data, C. H. Lin and colleagues [used a key generator (KG) and a chaotic map based on quan-tum mechanics, 10] to create a smart symmetric cryptography approach. It also involves generating random cipher codes, training an encryptor using grey relational analysis (GRA), and evaluating encrypted pictures. A. Shafique, J. Ahmed, M. U. Rehman, and M. M. Hazzazi et al. present a method for decryption that allows for the complete recovery of all original data (ref. [11] may be used to encrypt both text and pictures. Space-time encryption is used at the beginning and ending of the picture. The proposed method's middle section discusses frequency domain encryption using DWT.

Talha M. Abd El-Latif, AAbd-El-Atty B. [12] proposes a novel approach to encrypting medical images using quantum mechanics. The proposed method employs chaotic mapping and grey-coded data. The quantum image becomes distorted due to quantum grey coding. The quantum image is then encrypted using a key generator guided by the logistic-sine map in a quantum XOR operation. A NEQR quantum image representation was utilized to design the circuits for the proposed encryption and decryption method. The numerous different encryption algorithms now in use were analyzed by M. K. Hasan et al. [13], who considered execution time, memory consumption, and encryption quality. Medical image encryption was described by Y. Ding et al. [14], and a deep learning-based decryption method is proposed. The major educational network used for medical picture encryption and decryption is the Cycle-Generative Adversarial Network. Chen, Ping-Yuan, et al. [15] It is suggested to use a combination of two symmetric cryptography algorithms that share the same secret gold keys for the encryption and decryption of medical images. A controller based on optimization and a hash transformation using several secret gold keys are used to generate the relevant 2n-1pixel values included. In order to establish a complete framework for medical picture encryption, S. Ibrahim et al. [16] combined substitution boxes (S-boxes) with chaotic maps. A. Fekih, B. Vaseghi, S. Mobayen, and S. S. Hashemi. [17] To safeguard the transmission and/or storage of medical images, we provide a chaotic cryptosystem that employs coexisting chaotic systems as shared secret key generators. Through modeling and analysis, we test the practicability and efficacy of the suggested synchronization method.

A unique game-theoretical-based lossless medical image encryption system was suggested by J. Zhou, J. Li, and X. Di [18], with the aims of accurately identifying "the region of interest (ROI), avoiding ROI position information leak-age, and accomplishing leakage recovery of transform domain encryption." Drs. K. Shankar and Co.[19] As a solution to the problem of 1-dimensional chaotic cryptosystems' weak security and small key space, a security model was proposed. Extremely secure medical images with just a few subkeys are analyzed in this work using chaotic logistics and tent maps. Diffusion and confusion were included in the chaotic (C-function) procedure used to test the security. For both partial and total medical image encryption, Ravichandran D, Praveen Kumar P, Rayappan J, and Amirtharajan R presented a hybrid cryptographic approach in ref. [20] by making use of chaotic maps and deoxyribose. Through the simultaneous merging of many chaotic maps, the suggested approach produces very unpredictable encryption keys for use in medical imaging and color digital imaging. In order to encrypt and decode images, K. N. Hari Bhat [21] suggested key creation using a series of logistic maps and a chain of LFSR states. Protecting sensitive medical photos using watermarking and a shared encryption scheme was suggested by D. Bouslimi, G. Coatrieux, M. Cozic, and C. Roux [22]. P. Deshmukh, in ref. [23], suggests using the AES technique for image encryption and decryption to safeguard sensitive visual data from prying eyes. The AES symmetric key technique is one of the best available methods of encryption and decryption. With the help of MATLAB code, we may simulate an AES technique for picture encryption and decryption and then synthesize it.

In conclusion, several different algorithms and techniques [6]-[8], [17], [19], [24-30] were developed and introduced to improve data security; There are benefits and drawbacks to both options. It is common practice to evaluate and choose candidates based on characteristics such as time complexity, dynamic behavior, computation cost, attack resilience, and security. In this research, we propose a more stringent set of encryption criteria for the protection of clinical data, which might be included and used in the trustworthy e-healthcare tools shown in Figure 1. The primary goal of this research was to create a user-friendly data encryption system that used chaotic maps to achieve key sensitivity, low residual consistency, and acceptable high-satisfactory facts. Some recently suggested 2D transition methods result in illogical series matrices. It can be shown that these matrices have chaotic behavior by inspecting their bifurcation diagrams and Lyapunov exponents.

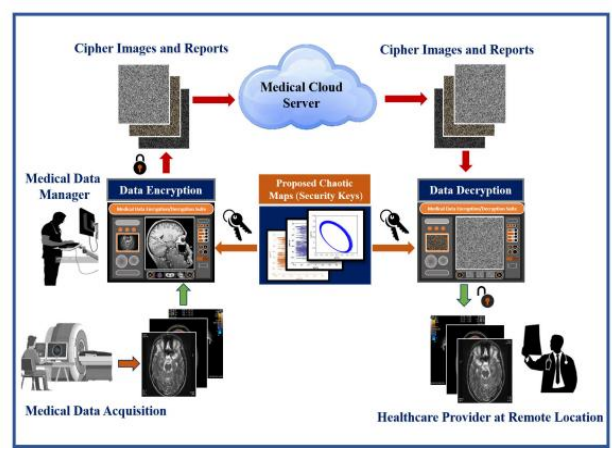

Figure 1. Cloud-Based Internet of Medical Systems with Security

The remaining portions of this paper are as follows. Section 3 depicts details on chaotic map characteristics, potential encryption frameworks, and medical image decoding. In Section 4 we describe both large-scale quantitative and qualitative investigations, as well as experiments conducted in computer simulation. Section 5 provides a wrap-up and closing thoughts.

### III. PROPOSED METHOD

*Chaotic Maps*

Because of its unique qualities and attributes, chaotic algorithms are finding increasing usage in cutting-edge cryptography studies, including as
a) The orbital evolution is very sensitive to the initial circumstances and the control settings, and it is also highly unpredictable.
b) Strong encryption achieved by combining a rigorous and
c) Putting in place software is a breeze.

Two-dimensional chaotic maps were employed in our study to provide safe approaches to medical imaging and data encryption. The suggested framework is nonlinear and discrete-time, and it is characterized by dynamically chaotic behavior. Unpredictability and ergodicity properties allow searches to be conducted more rapidly than those based on probability distributions [5, 15], 31]. For the purpose of medical picture encryption, two different maps are used and additional possible maps are investigated.

A chaotic map is a function that, by definition, exhibits chaotic behavior. A given point that remaps (xn, yn) is defined by the mathematically controlled chaos of the proposed maps.

$(x_{n+1}, y_{n+1})$:

$$y_{n+1} = y_n - r * \tanh x_n$$
$$x_{n+1} = \sin x_n + \cos y_n \quad (1)$$

$$y_{n+1} = b * x_n^2$$
$$x_{n+1} = x_n + y_n^2 - a * r \quad (2)$$

In the first proposed map (1), we see a fresh 2D economic model [7]. This term is used to describe a novel nonlinear dynamic model in which the mapping of (xn, yn), where r is a parameter that controls how many times an iteration is performed (n). The second suggested map (2) is a new take on the classic logistic map; it has three control variables (a, b, and r), but displays chaotic behavior like the Henon map. To construct such a map, we first need a model, as in (1) or (2), and some beginning values for the parameters. Next, we arbitrarily create the map's x0 and y0 starting points. The last step involves repeating an iterative process n times. $x_{n+1}$ and $y_{n+1}$.

Additionally, the dynamic behaviors of the map display both jerky motion and irregular orbits. The proposed technique iteratively transforms the original 2D picture into a 1D representation, which is then encoded.

In this context, the suggested chaotic maps are put to good use. Pixels may be moved about with one, and their densities adjusted with the other. The method employs several logical processes that, taken as a whole, carry out the mergers and acquisitions necessary for encryption, making it both more secure and quicker.

*Proposed Encryption Scheme and Decryption Scheme:*

We provide a new method for encrypting and decrypting images that is based on chaotic processes. At its heart, the suggested solution is an image-splitting process that divides the input picture's pixels into many portions (Is, s2) before encryption [7]. We use s = 2 since our two unique keymaps may encode either the left or right side of the image. Multiple splits and multiple maps, or combinations of maps, may be used to encode distinct parts of a picture.

A large enough number of Is sets, the reasoning goes, will help with the regeneration of I, even while individual picture parts or splits carry no meaningful information. Although this approach is effective, the recovered picture may be of poor quality owing to the loss of color and contrast. A novel and speedy method is shown that utilizes chaotic keys, which not only help circumvent this limitation but also reduce the likelihood of mistake throughout the encryption and decryption processes, keeping the quality of the recovered picture intact. The proposed methods are attack-resistant and need little time and bandwidth to implement. To completely encrypt a picture, two steps of confusion (also called permutation) and diffusion [16] methods are employed. Other methods are geared at using chaotic states and basic picture data to randomly shuffle and replace pixel values, resulting in a cipher image that seems like random noise.

*Encryption Process*

Figure 2 shows a flowchart of our chaotic medical photo encryption method. An input plain image (Dimensions by Height by Width) is used to generate a 1D matrix where the pixel values are added together in columns. The image's height and width are denoted by H and W, respectively. where D is the picture depth, which is 1 for both monochrome and color images. 3.

The processed 1D matrix has two portions of size (DHW)/2, one for (P1) and the other for (P2). Separated parts of the code are encrypted using the two proposed maps. "Recommended mappings are (1) and (2), where r, a, and b are external parameters that influence the chaotic behavior of the map and act as keys in the encryption process." Individual image partitions may be encoded using a variety of maps, either alone or in combination. The fragmented pictures are encrypted using XOR and random key sequences.

*Permutation Process*

Chaotic maps are used to create turbulent groups through the permutation (or confusion) process. In order to reassemble the image into its scrambled form, the original picture is used as a seed to construct a record-keeping network out of the chaotic sequences. The disconnection between adjacent pixels that occurs during the ordering process makes full re-identification of the original image impossible [27]. Thus, the dispersal method enhanced security. One of the factors that decides whether a cipher pixel is altered is the value of the pixel that was modified before it. Therefore, a modification to even a single pixel may have far-reaching consequences for the rest of the pixels in the image. The resulting 1D splitter matrix replacements are then XORed into the key sequence. Three times, a different random key sequence is used to perform the permutation process.

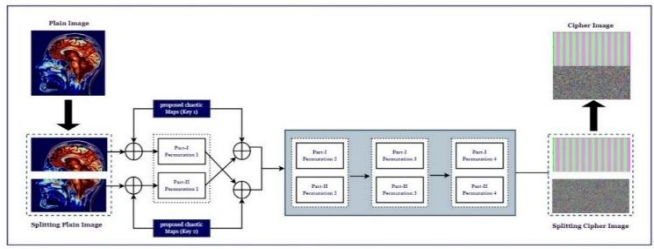

Figure 2. Flow Diagram Illustrating the Encryption Process for the Proposed Scheme

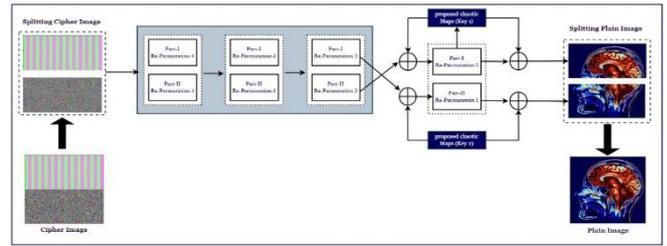

Figure 3. Flow Diagram Illustrating the Decryption Process for the Proposed Scheme

**Algorithm 1** Proposed Algorithm for Encryption Process"

1. The original image P (size H W) is input, and the encrypted, noisy image C is produced.
2. Convert P to a 1D pixel array, then split it into two vectors, P1 and P2, of sizes H W/2.
3. In order to generate the chaotic sequence, we apply equations (1) and (2) n times (where n is the length of the array).
4. Replace Step 3 with the recommended chaotic maps; the corresponding condition is in Equation (3). Be mindful that all grayscale components are within the range [0, 255].

"$P(t) = \mod(\text{round}(10^{12} P(t)), 256)$, where $1 \leq t \leq H \times W/2$. (3)

5. Using the XOR operation in Equation (4), generate diffused vector.

$D_i = P_i \oplus K_i$ ($\oplus$ is the bit-wise XOR) (4)

6. Permute the sequence of $P_1$ with the 1st vector of the 1st proposed map and $P_2$ with the 1st vector of the 2nd proposed map in Equation (5).

$p_i = D_i(K_i)$ (5)

7. Swap the splits and again generate diffused vector Equation (4).
8. Again Re-Permute (3 times) the sequence of $P_1$ with the 2nd vector of the 1st proposed map and $P_2$ with the 2nd vector of the 2nd proposed map in Equation (5).
9. Join the two encrypted parts to get the cipher picture, and create the final matrix with Equation (6).

$C = \text{reshape}(p_i, H, W)$" (6)

*Decryption Process"*

Decryption works by doing the opposite of what happens during encryption. A simplified diagram of our decipherment procedure is shown in Figure 3. The same mysterious keys may be used to generate both disordered vectors and chaotic record sets throughout the encryption process. The first steps in decryption include the use of inverse diffusion, confusion, and combination.

A diffuse inverted vector is generated in the center of a 2D matrix with the dimensions H W, which is originally translated from the encoded picture C. In addition, we separate the common alternate vector (P1p, P2p) into its component parts. Finally, the inverse levels (P1, P2) for each partial region were generated utilizing the chaotic recording setup and a variety of techniques.

**Algorithm 2** Proposed Algorithm for Decryption Process

1. C, a noisy encrypted picture, serves as input, while P, a clean version of C, serves as output.
2. Convert C to a 1D pixel array, and then split it into two H W/2 sized vectors, C1 and C2.
3. Using Equation (7), invert and permute the vector three times.

$D_i(K_i) = c_i$ (7)

4. Generate the de-shuffled vector using Equation (8)

$C = D \oplus K$ (bit-wise XOR operation $\oplus$) (8)

5. "If you take the second chaotic sequence and combine it with the first advised map to scramble C1 and the second proposed map to scramble C2, you may get an inverse permuted vector with C=K (index).
6. To generate an anticlockwise permutation of C1 and C2 sequences, use the first vectors from the first suggested map and the second vectors from the second recommended map, respectively, and invert the splits, as shown in Equation (7)."
7. Equation (8) may be used to re-create the unscrambled vector.
8. Using the chaotic index sequence and reshaping vector components, make C1 and C2.
9. Combine resultant halves, $C_1$, and $C_2$, to get P.

IV. EXPERIMENTAL RESULTS

Using the proposed maps, an encryption/decryption technique has been constructed and tested on a number of medical MRI scans (including brain). Figure 4 shows the final encrypted and decrypted data.

The simulation platform was built on an Intel(R) Core(TM) i5-24004 computer with 8 GB of RAM and a 512 GB solid-state drive. All simulation analyses were performed in Python on the Google Colab. All of the simulation experiments have been repeated indefinitely.

Figure 4(a) shows the unencrypted medical picture (MRI scan) while Figure 4(b) depicts the encrypted version. Figure 4(c) and Figure 4(d) depict the encrypted and decrypted versions of the original medical picture (MRI scan), respectively. The original medical picture, after being encrypted, is shown in Figure 4(e).

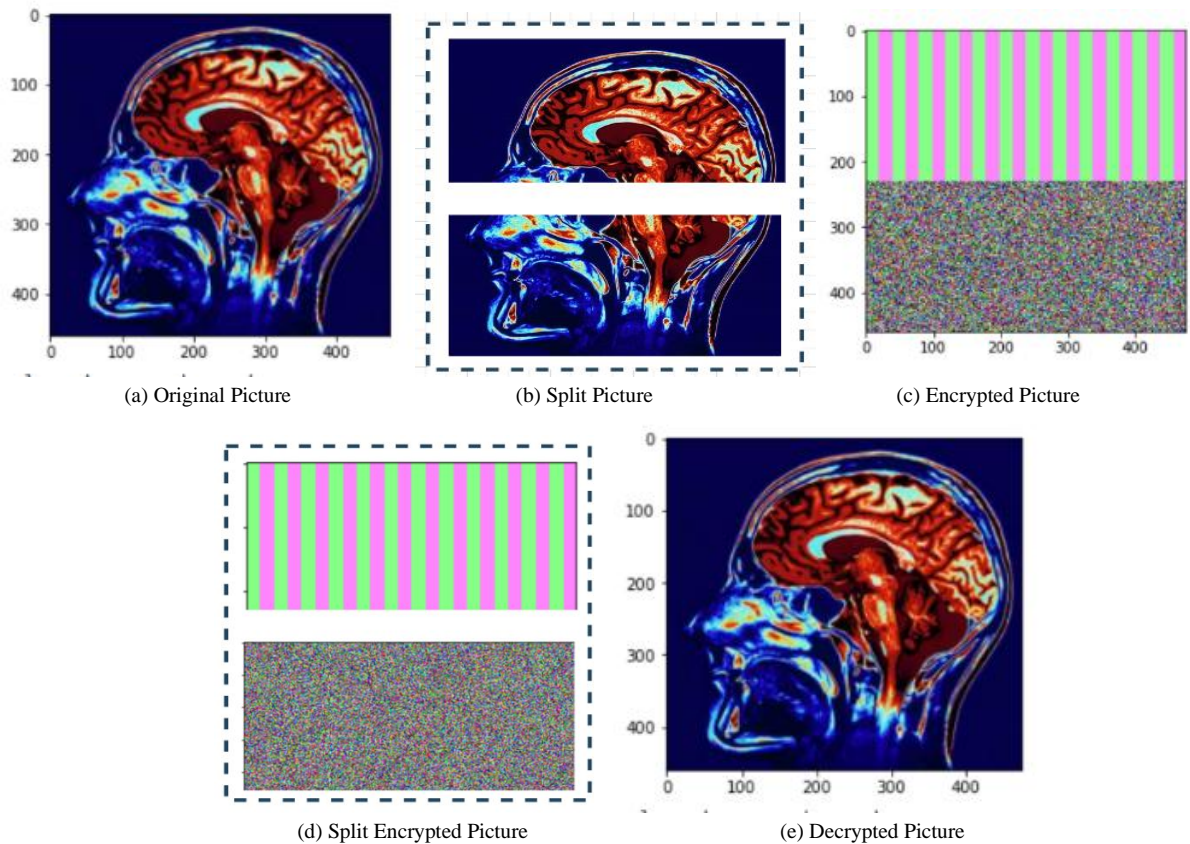

(a) Original Picture  (b) Split Picture  (c) Encrypted Picture

(d) Split Encrypted Picture  (e) Decrypted Picture

Figure 4. Encrypted and Decrypted images for an MR-Brain Picture

Since the suggested maps produce disordered results when used with varied combinations of the parameters, it is important to choose values for the parameters that will serve as controls. For our simulated experiments, we used the values of r = 17.0 for (1) and 2.35 for (2). "The r-values for both maps were selected at the points shown in Figure 4, which are diagrams of bifurcation and Lyapunov exponents. The values for the two variables in Eq. (2), a and b, were derived from a previous study that examined map trajectories in a simulation experiment with varied values of a and b [0.1: 0.1: 1]. In a chaotic setting, the optimal values for a and b on the map are [0.5, 0.3]. Although these values are totally subjective, we have settled on 0.1 for the initial two map coordinates, $x_0$ and $y_0$.

*Bifurcation and Lyapunov exponent*

Lyapunov and bifurcation diagrams are used to examine the potentially chaotic character of the suggested maps. A bifurcation diagram shows the possible outcomes of a recursive system. In our scenario, as seen in the figure below Figure 5(a), the system splits in half at the number 1.5, again at the number 2, and so on, until it reaches infinity by repeatedly doubling itself. The diagrams show how disorder may spread. Close inspection of this bifurcation diagram reveals sparse regions where values are scarce. More information on the bifurcation diagrams for the first and second maps is provided in Figure 5(a) and Figure 5(b), respectively. The Lyapunov exponents characterize how sensitive a system is to perturbations in its starting conditions and predictability. If the exponent is positive, the system will display chaotic behavior regardless of where it is initially set. Lyapunov exponents for the first and second maps are shown in Figure 6(a) and (b), respectively. Bifurcation diagrams are examined by systematically altering a single variable while holding all others constant. Bifurcation diagrams for a and b are investigated. Figure 5 depicts the bifurcation diagram of the chaotic system with a variable parameter r, while the other two parameters, a and b, are held constant at values of 0.1 and 0.1, respectively.

*Histogram Analysis*

Histogram analysis, a common statistical method, may be used to evaluate an image encryption system qualitatively. One of the simplest ways to display the picture encryption quality is via the histogram analysis given in Figure 7. The goal of most reliable picture encryption methods is to make the original image completely unrecognizable. Therefore, an ideal picture encryption method will provide a cipher image with a histogram of intensities that is normally distributed. Figures 7(a), (b), and (c) show the histogram analysis for the decrypted and unencrypted versions of the selected example medical image (MRI-Brain). Figure 7(d), Figure 7(e), and Figure 7(f) show that the histograms of the cipher photographs are evenly distributed over the range of grey values, but the histograms of the original images are clustered around particular grey values. This exemplifies the security provided by encrypted data against statistical assaults.

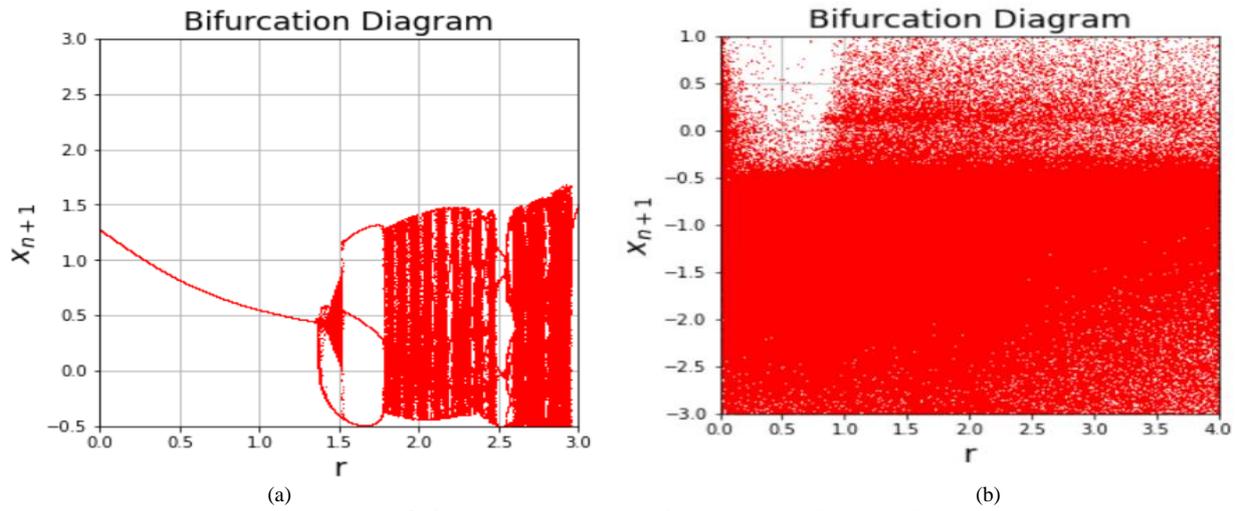

(a)        (b)

Figure 5. Bifurcation diagram of (a) chaotic map1 (b) chaotic map2

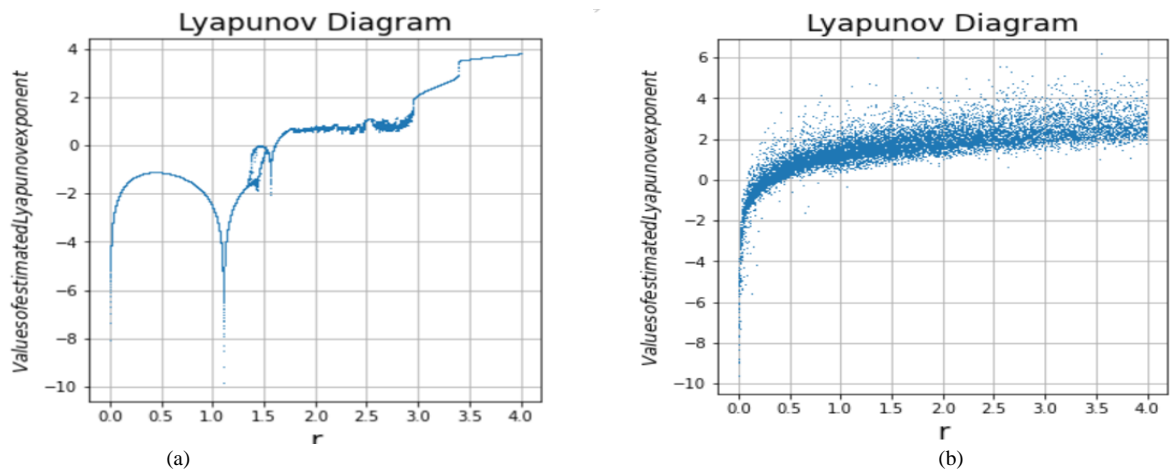

(a)        (b)

Figure 6. Lyapunov diagram of (a) chaotic map1 (b) chaotic map2

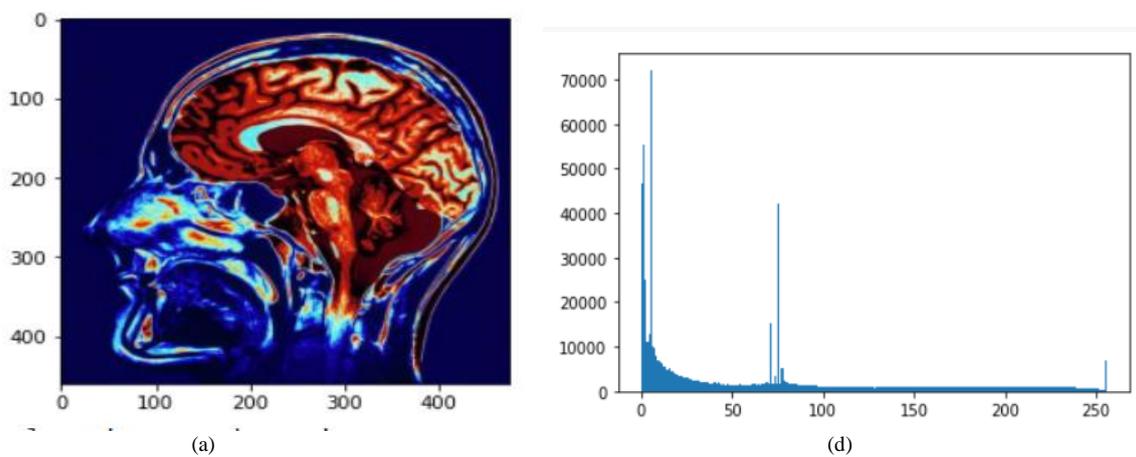

(a)        (d)

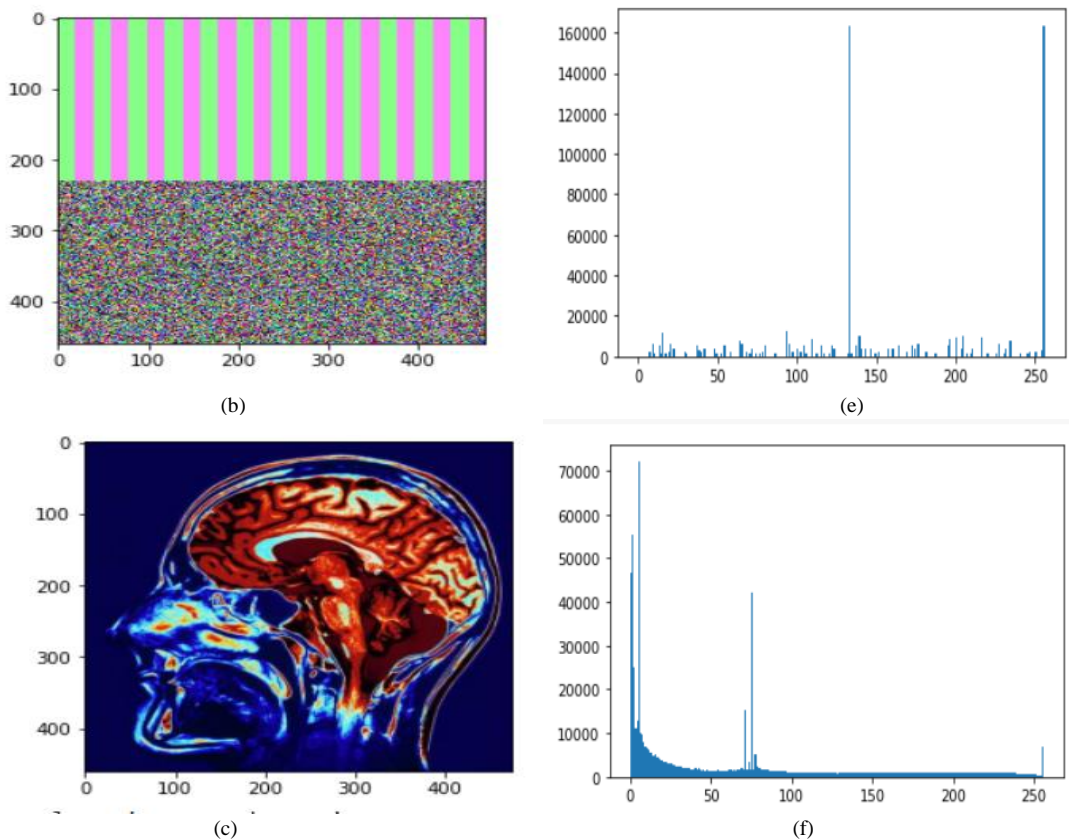

Figure 7. Qualitative findings of the proposed structure using MRI-Brain image - original images, histograms

*MSE and PSNR of the Proposed System"*

In fact, there are a number of indications that may be used to qualitatively assess the resilience against differential assaults. The effectiveness of a cryptosystem may be evaluated in this way.

One such statistic used for assessment is the Mean Square Error (MSE). The Mean Squared Error (MSE) calculates the distance between two pictures. Let I1 and I2 be the unmodified and encrypted pictures, respectively, and let m and n be the image's dimensions, Equation (9) provides the formula for MSE. –

$$\text{MSE} = \sum_{m,n} [I_1(i,j) - I_2(i,j)]^2 / (m \times n) \qquad (9)$$

One such statistic used for assessment is the Mean Square Error (MSE). The Mean Squared Error (MSE) calculates the distance between two pictures. Let I1 and I2 be the unmodified and encrypted pictures, respectively, and let m and n be the image's dimensions, Equation (9) provides the formula for MSE. –

"PSNR= $10 \log_{10}(R^2/\text{MSE})$ where R=255        (10)

The results of a qualitative comparison of the encrypted and decrypted images are shown in Table 1.

TABLE 1. Result of MSE and PSNR.

| | |
|---|---|
| MSE (4 permutations) of encrypted image: | 20511.97642481502 |
| PSNR (4 permutations) of encrypted image: | 5.010728521798612 |
| MSE (4 permutations) of decrypted image: | 0.0 |
| PSNR(4 permutations) of decrypted image: | inf" |

The PSNR that I measured was 5.02 dB MSE, and the MSE was 20511.98. The PSNR of the encrypted version is close to that of the original, but the PSNR of the original is much lower. This leads us to believe that the cryptosystem is trustworthy.

*Comparison*

If we insert a few more permutations in the middle of the procedure, we can see what happens to the MSE and PSNR in Tables 2 and 3, respectively. Graphically depicting the accumulating impact of the encryption effect are scatter plots of MSE and PSNR data with 6 rounds of permutations (Figure 8 and Figure 9, respectively). We find that after 4 iterations of permutation, the PSNR and MSE values are significantly altered, with lower MSE and higher PSNR than the previous iteration.

TABLE 2. Result of MSE

| No. of Permutations | MSE of Encrypted Image |
|---|---|
| 1 | 15080.58965 |
| 2 | 13670.55135 |
| 3 | 17296.24374 |
| 4 | 20511.97642 |
| 5 | 20511.97642 |
| 6 | 20175.71965 |

TABLE 3: Result of PSNR

| No. of Permutations | PSNR of Decrypted Image |
|---|---|
| 1 | 6.34662038 |
| 2 | 6.772943304 |
| 3 | 5.751285641 |
| 4 | 5.010728522 |
| 5 | 5.010728522 |
| 6 | 5.082513262 |

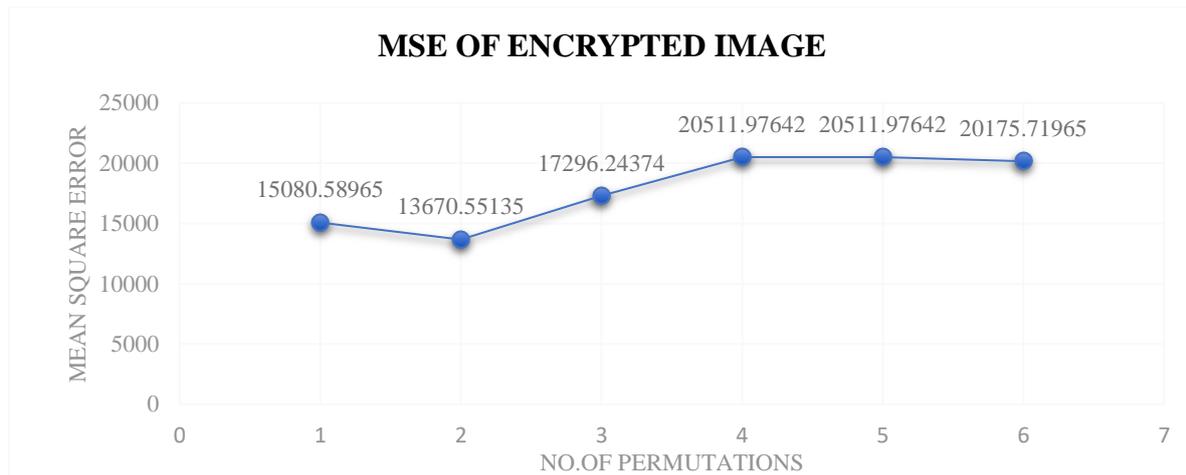

Figure 8. The Distribution of MSE with the number of permutations

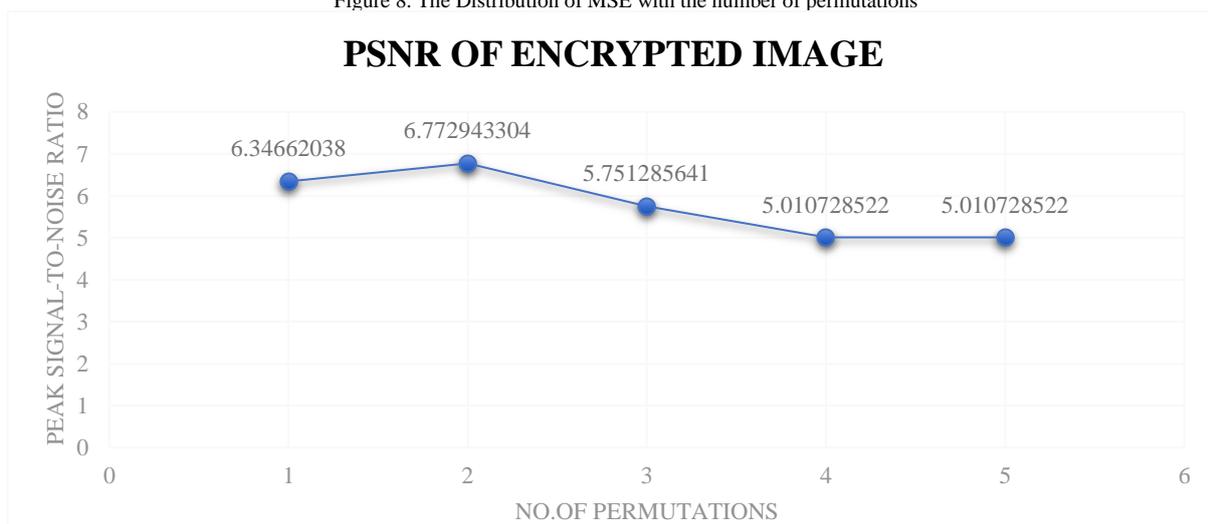

Figure 9. The Distribution of PSNR with the number of permutations

Conclusion The 4-permutation encryption technique yields the best MSE and PSNR results.

## V. CONCLUSION

In this research, we lay out a fresh and secure method for securing medical picture storage in the cloud for potential usage in IoH systems. In our experiments with two novel chaotic maps, we found that they are very sensitive to the initial seed circumstances, exhibit highly unpredictable behavior, and exhibit severe chaotic behaviors. Since they are based on dynamic "analysis and validation using a bifurcation diagram and the Lyapunov exponent," the proposed maps are often hyperchaotic, very sensitive, and extremely complex."

In addition, the suggested pipeline technique (1) employs disarray and distribution structures with three modifications, unlike one-time-key encryption systems, and (2) provides for extra information limits apart from the initial image and the mystery key. The latter has the advantage of letting you change encrypted data values without compromising the security of your private keys. Therefore, our approach avoids the shortcoming of one-time-key methods while providing various advantages, such as the ability to quickly and securely encrypt a large number of images with a single key and improved encryption quality, resilience, and speed.

This has been shown using data from a number of studies as well as test medical photos. Figure 8 displays the calculated MSE and PSNR values of 20175.72 and 5.09 dB, respectively. Finally, the proposed framework has the potential to increase the safety of patient information and improve the aesthetic quality of medical images. It cannot be overstated how useful the proposed pipeline would be for any multimedia encryption application outside of the medical field.

*Future Scope:*

Hyper- and high-dimensional chaos maps, which possess superior chaotic properties, should be the focus of future research. In order to build composite hyper-chaotic dynamical systems and composite high-dimensional chaotic systems, one may use the remarkable structure of composite discrete chaotic systems to increase the system's complexity and produce exceptional chaos. However, we need to think about how much time and effort it will take to use these systems for picture encryption.

## Authors' Profiles

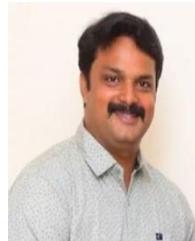

"Dr. Chandra Sekhar Sanaboina is an Assistant Professor in Computer Science and Engineering department currently employed by University College of Engineering Kakinada, JNTUK University, Kakinada. From Koneru Lakshmaiah College of Engineering, he graduated with a Bachelor of Technology (B. Tech) in Electronics and Computer Science Engineering in 2005." At Vellore Institute of Technology, he then earned a Master of Technology (M. Tech) in Computer Science and Engineering in 2008. In 2020 completed his Doctor of Philosophy (Ph. D) at JNTUK in the field of the Internet of Things. He had been a teacher for more than 15 years and a researcher for about 12 years. His research interests include Artificial Intelligence, Data Science, Cryptography, Machine Learning, the Internet of Things, and Wireless Sensor Networks.